# Oxide Layer Thickness Effects on the Resistance Switching Characteristics of Ti/TiO$_2$-NT/Au Structure

Irina B. Dorosheva
*NANOTECH Center*
*Ural Federal University*
Yekaterinburg, Russia
i.b.dorosheva@urfu.ru

Alexander S. Vokhmintsev
*NANOTECH Center*
*Ural Federal University*
Yekaterinburg, Russia
a.s.vokhmintsev@urfu.ru

Robert V. Kamalov
*NANOTECH Center*
*Ural Federal University*
Yekaterinburg, Russia
rvkamalov@urfu.ru

Aleksey O. Gryaznov
*NANOTECH Center*
*Ural Federal University*
Yekaterinburg, Russia
aogryaznov@gmail.com

Ilya A. Weinstein
*NANOTECH Center*
*Ural Federal University*
Yekaterinburg, Russia
i.a.weinstein@urfu.ru

*Abstract* — Self-ordered nanotubular titania TiO$_2$-NT with outer tube diameter of 45 nm are synthesized using the anodic oxidation of titanium foil. Four sets of memristors with 100 μm diameter based on Ti/TiO$_2$-NT/Au sandwich structures with an oxide layer thickness of 80, 120, 160 and 200 nm are fabricated. Current-voltage (CV) characteristics for the obtained samples in the static and dynamic operation modes are studied. Resistance in high and low resistance states is estimated. Basing on the analysis of the CV characteristics in dynamic mode (> 14 000 switchings) a prospective of use for synthesized Ti/TiO2-NT/Au micromemristors with oxide layer thickness of 160 nm in non-volatile memory is shown.

*Keywords—anodic titania, titanium dioxide nanotubes, nanotubular structure, memristor, resistive switching*

## I. Introduction

At present much attention is paid to studying the processes of resistive switching in thin layers of semiconductors and dielectrics to create new computing devices for nanoelectronics and integrated neuromorphic networks [1, 2]. For these applications, titanium dioxide is actively investigated since 2008 year, when on the its basis a passive memristor element was produced for the first time [3].

Memristive behavior in the metal/oxide/metal structures is generally caused by the movement of anionic, cationic vacancies, etc. in the dielectric layer at an external electric field strength >10$^6$ V/m. It is known [1, 3] that thickness and defectiveness of the oxide layer defines the ratio between the electrical parameters of the structure in low ($R_{LRS}$) and high resistance ($R_{HRS}$) states, determing the noise immunity. Previously, it was shown [4–7], that memristive effect is not only founded in compact oxide layers but also in self-ordered arrays of titania nanotubes TiO$_2$-NT, obtained by the electrochemical oxidation (anodization) of titanium. It is also known [8] that the parameters of anodization essentially influence on the morphology and defectiveness of the oxide layer. Thus, the adjustment of the synthesis parameters to fabricate the memristor elements and memory devices with the optimal specification and performance is a crucial application task of the nanoelectronics. Thereby, the aim of this research is to study the influence of the oxide layer thickness on the parameters of static and dynamic modes of Ti/TiO$_2$-NT/Au micromemristor cells switching.

## II. Experimental

In the present work, the anodization method of Ti foil and sputter deposition of Au top contacts was used to obtain the memristive structure. The anodization of Ti substrate (VT1-0 type, Russia) was carried out in the electrochemical cell while maintaining constant temperature of 20 °C. M1, M2, M3 and M4 semples of memristors were formed at constant voltage of 10 V and anodization time of 5, 10, 15 and 20 min, respectively. The solution of ethylene glycol 45 vol. %, glycerol 45 vol. % and water 5 vol. % with addition of ammonium fluoride 1 wt.% was used as an electrolyte. All chemicals were analytical grade purity.

The samples morphology was studied using SIGMA VP Carl Zeiss scanning electron microscope (SEM) in high vacuum mode using the InLens detector.

All synthesized samples of TiO$_2$-NT were covered with golden top contacts by the masking method (see Fig. 1) using magnetron sputtering unit Q150T ES Quorum Technologies. More than 100 memristors for each sample were fabricated in a single process. Figure 1 shows typical Au contacts with a diameter of 100 μm and a thickness of 100 nm. Image is obtained with Axio CSM 700 confocal optical microscope.

Current-voltage (CV) measurements were carried out using PXI-4143 National Instruments programmable power supply and MPS 150 Cascade Microtech microprobe station. In the static mode bias harmonic voltage with frequency of 0.01 Hz and amplitude 1–2 V was applied to top electrodes. Ti electrode was grounded. The switching procedure for each memristor was performed with a positive voltage of up to 3 V before the current-voltage measurements. The current through the sandwich structure was limited to 5 mA (M1 and M4) and 10 mA (M2 and M3).

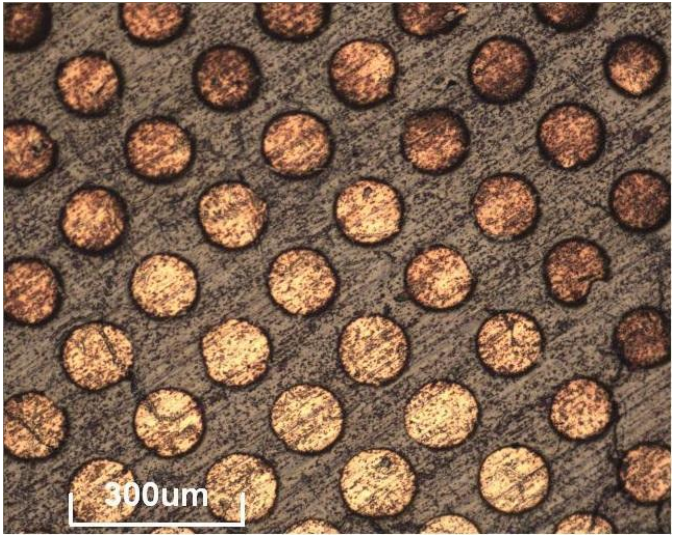

Fig. 1. A typical external view and configuration of Au microcontacts on the anodic titania surface

The measuring installation and the methods for recording the CV curves in static and dynamic modes are described in detail in [7].

### III. RESULT AND DISCUSSION

#### A. Characterization of TiO$_2$-NT layer

As an example, Fig. 2 shows the SEM images of the surface of anodic titania (M4 sample). As follows from the images, the 20 min synthesis in the above described conditions leads to the formation of the TiO$_2$ nanotubular structure with the outer diameter of nanotubes d = 45 ± 10 nm on the titanium surface. The thickness of the TiO$_2$ layer grows with increase of the anodization time. It is determined that the thickness of the oxide layer is equal to h = 80, 120, 160 and 200 nm for samples M1, M2, M3 and M4, respectively.

#### B. The current-voltage measurements under static mode

CV characteristics, obtained for all samples, indicate the presence of the memristor behavior. Fig. 3 shows the CV curves for three complete switching cycles of the studied memristor structures. It should be noted that initially after the fabrication the memristor samples are in the HRS state. For correct operation, in accordance with the logic below, it is required to perform the electro activation (see Fig. 3, red line and Experimental) to switch the structures to the LRS state.

Fig. 3 shows that the decrease of the bias voltage U leads to the switch of the memristors to the HRS state at U = 0.5–1.5 V negative voltage. Following increase of U leads to the LRS state switch at a positive voltage of U = 0.5–1.5 V. Thus, the bipolar resistive switching is registered for all fabricated

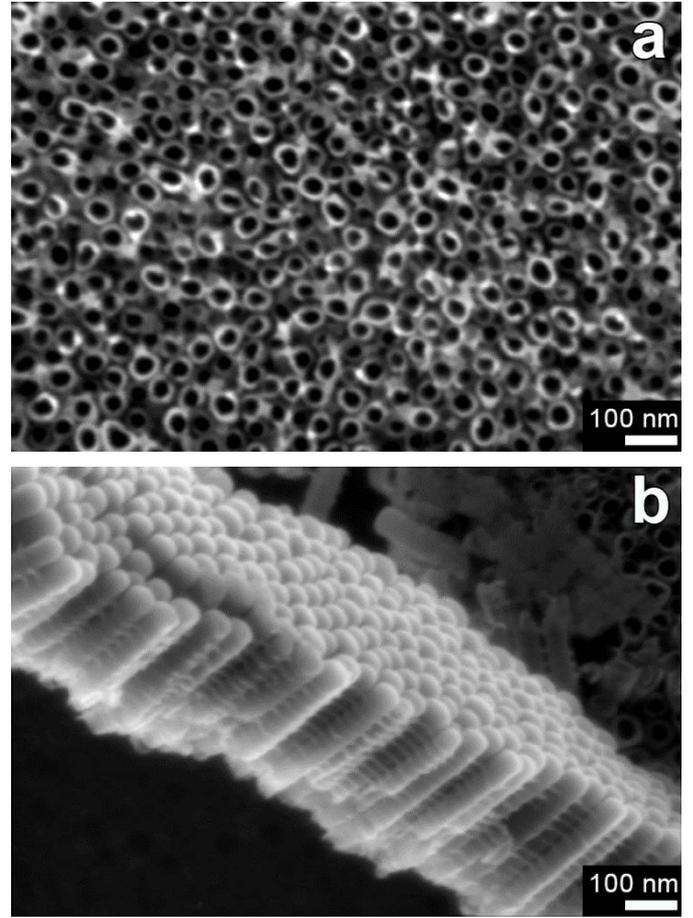

Fig. 2. Titanium dioxide nanotubular structure of the M4 sample: a) top view; b) cross-sectional view.

Ti/TiO$_2$-NT/Au sandwich structures. This fact agrees well with the reports [4–7]. By contrast, it can be noted that the values of the current through the Ti/TiO$_2$-NT/Au structure are more than 5 times lower than [4, 6].

Analysis of the CV characteristics for M1–M4 samples demonstrates that for thicker TiO$_2$ layers higher value of the applied voltage is required to switch the structure from the HRS to the LRS and vice versa. The number of switching cycles for M2, M3 and for M1, M4 is N ≈ 20 and N ≈ 5, respectively.

For all CV curves the average values of the resistance of the memristor sets in high (<$R_{HRS}$>) and low-resistance (<$R_{LRS}$>) states were estimated at the same voltage of U = ± 0.2 V (see Table 1). Absolute error in determining of <$R_{HRS}$> and <$R_{LRS}$> is equal to 10 %. The best ratio <$R_{HRS}$> / <$R_{LRS}$> = 186–134 is obtained for the M3 set of memristors with an oxide layer thickness of 160 nm. As opposite to M1 and M4 samples, where the properties of the resistive switching are degraded already by the fifth switching cycle, the CV characteristics for the M3 sample are relatively stable, reproducible at N ≥ 20 cycles and possessing the most symmetrical form.

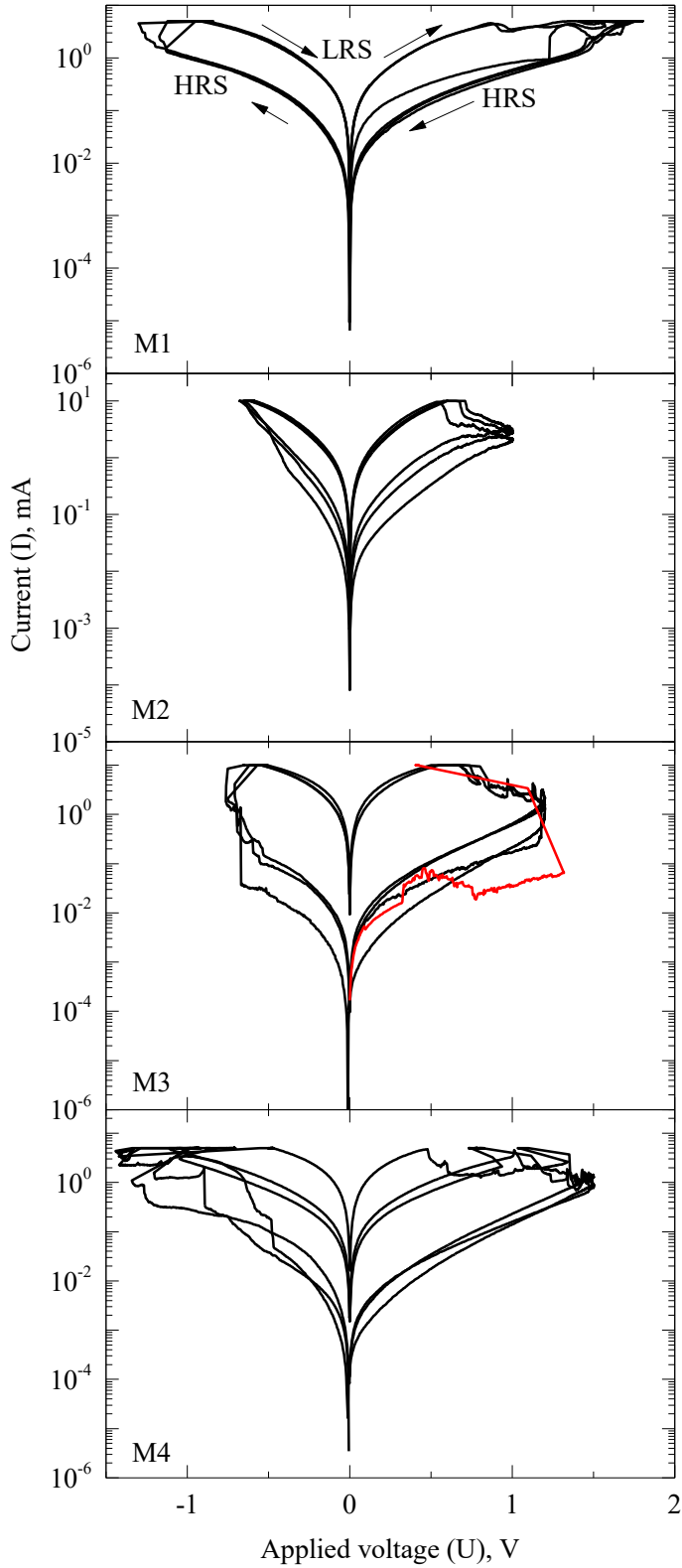

Fig. 3. CV curves of Ti/TiO$_2$-NT/Au structures. Black lines: full cycles of resistive switching. Red line: the example of the electro activation curve for the M3 structure. The arrows show the order of the CV record. Sample is given in bottom left corner of the figure.

TABLE I. ELECTROPHYSICAL PARAMETERS OF Ti/TiO$_2$-NT/Au SANDWICH STRUCTURES

| Parameter | Samples | | | |
|---|---|---|---|---|
| | M1 | M2 | M3 | M4 |
| h, nm | 80 | 120 | 160 | 200 |
| <$R_{HRS}$>, kOhm | 4.7 | 26.7 | 11.4 | 32.7 |
| <$R_{LRS}$>, Ohm | 505 | 61 | 85 | 230 |
| $R_{HRS}$ / $R_{LRS}$ | 9 | 441 | 134 | 142 |
| N | ≤ 5 | ≤ 20 | ≥ 20 | ≤ 5 |

*C. The current-voltage measurements in the dynamic mode*

Fig. 4 shows the dependencies of electrical resistance of the M3 memristor as a function of the number of switchings (n). It can be seen that for the given switching parameters in dynamic operation mode (see Experimental), the $R_{HRS}$ increases from ≈ 7 to ≈ 30 MOhm, the $R_{LRS}$ decreases from ≈ 1 to ≈ 0.1 MOhm, and the $R_{HRS}$ / $R_{LRS}$ ratio changes from ≈ 7 to ≈ 30 with increasing the value of n.

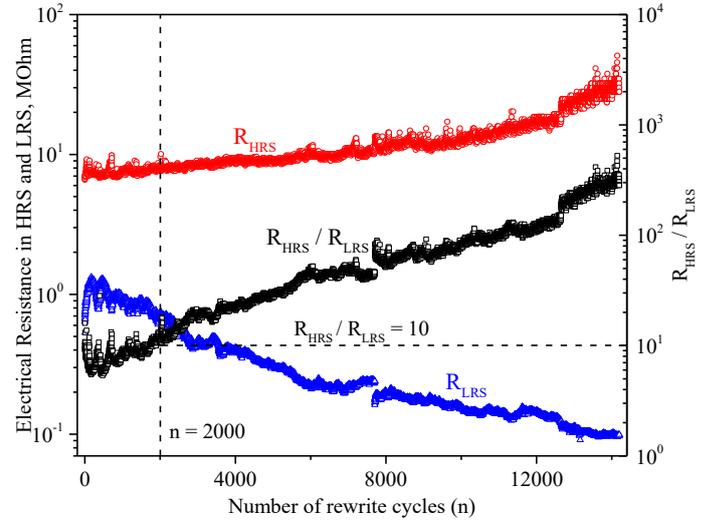

Fig. 4. The dependencies of electrical resistance $R_{HRS}$, $R_{LRS}$ and their ratio from the number of switching cycles for M3 memristor.

The obtained high values of the resistance of the M3 sample for the dynamic mode in comparison with the estimates of similar parameters from the static CV characteristics (see Table 1) indicate the use of non-optimal switch conditions for the studied memristor structure. The selection of switching conditions for memristors in dynamic mode of operation is an important stage that stands beyond the scope of this study. It should be noted that for n > 2000 the ratio of $R_{HRS}/R_{LRS}$ > 10. This fact indicates the possibility of practical application of Ti/TiO$_2$-NT/Au structures with an anodized oxide layer thickness of 160 nm as elements of memristive memory. Thus, it can be concluded that the method described in the work is promising as a technology for the synthesis of real functional media for elements of non-volatile memory.

## IV. CONCLUSION

The synthesis of TiO$_2$-NT nanotubular structures was carried out by anodizing titanium foils with a process time varying from 5 to 20 min in a fluorine-containing solution of electrolyte at a voltage of 10 V. It is shown that the obtained oxide layer has a nanotubular structure with an outer tube diameter of 45 ± 10 nm and its thickness varies in range of 80-200 nm.

The CV characteristics of the Ti/TiO$_2$-NT/Au micromemristors structures in the full cycles of resistive switching (static mode) and simulating the processes of multiple switching (dynamic mode) are investigated. The influence of thickness on the resistances ratio in the high- ($R_{HRS}$) and low-resistance ($R_{LRS}$) states and on the number of switching cycles N is analyzed. The lowest value of $R_{HRS}/R_{LRS}$ = 9 and a small number of N ≤ 5 are observed at an 80 nm oxide layer thickness. The ratio of $R_{HRS}/R_{LRS}$ > 100 with N ≥ 20 and N ≤ 5 were obtained for samples with an 160 and 200 nm oxide layer thickness, respectively. The best value (from practical point of view) of $R_{HRS}/R_{LRS}$ > 400 and N ≤ 20 were found for 120 nm thickness structure. The operability of micromemristors with 160 nm thickness of TiO$_2$-NT functional layer for more than 14 000 switchings is shown.

Basing on the analysis of the CV characteristics (symmetry, number of switching cycles, $R_{HRS}/R_{LRS}$ ratio) in static and dynamic operation modes, it is discussed the Ti/TiO2-NT/Au sandwich structures are promising material for the fabrication of memristor memory elements.

## V. ACKNOWLEDGMENTS

A.S.V. and I.A.W. thank Minobrnauki initiative research project № 16.5186.2017/8.9 for financial support. Contribution to the study from R.V.K. was funded by the RFBR according to the research project № 18-33-01072. It was supported by Act 211 Government of the Russian Federation, contract № 02.A03.21.0006.